\documentclass[pra,twocolumn,nofootinbib,eqsecnum,floatfix,superscriptaddress]{revtex4}
\usepackage{epsfig}
\usepackage{bm}% bold math
\usepackage{amsmath}
\input epsf
\numberwithin{equation}{section}
\renewcommand{\theequation}{\arabic{section}.\arabic{equation}}

\def\psih{\hat\psi}
\def\Hc{{\cal H}}

\def\be{\begin{equation}}
\def\ee{\end{equation}}

\begin{document}

\title{The Quantum Mechanical Arrows of Time\footnote{Talk at the conference to honor Yakir Aharonov's 80th Birthday,
{\it Fundamental Aspects of Quantum Theory: A Two-Time Winner},
Chapman University, August 15-18, 2012.}}

\author{James B.~Hartle}

\email{hartle@physics.ucsb.edu}

\affiliation{Santa Fe Institute, Santa Fe, NM 87501}
\affiliation{Department of Physics, University of California,
 Santa Barbara, CA 93106-9530}

\date{\today }

\begin{abstract}
The familiar textbook quantum mechanics of laboratory measurements incorporates a quantum mechanical arrow of time --- the direction in time in which state vector reduction operates. This arrow is usually assumed to coincide with the direction of the thermodynamic arrow of the quasiclassical realm of everyday experience. But in the more general context of cosmology we seek an explanation of all observed arrows, and the relations between them, in terms of the conditions that specify our particular universe. This paper investigates quantum mechanical and thermodynamic arrows in a time-neutral formulation of quantum mechanics for a number of  model cosmologies in fixed background spacetimes. We find that a general universe may not have well defined arrows of either kind. When arrows are emergent they need not point in the same direction over the whole of spacetime. Rather they may be local, pointing in different directions in different spacetime regions. Local arrows can therefore be consistent with global time symmetry. 
\end{abstract}

%\centerline{\Huge  AOT2 \today}

\maketitle

\section{Introduction}

In his 1932 book \cite{vNeu32} von Neumann summarized the quantum mechanics of a subsystem of the universe that is sometimes measured but otherwise isolated.  Two laws of evolution for the quantum state of the subsystem were postulated.  The first is the Schr\"odinger equation that specifies how the state evolves  in time when the subsystem is isolated: 
\be
\label{seqn}
i\hbar \frac{d|\psih(t)\rangle}{d t} = H |\psih(t)\rangle  \quad \  \text{(I)} \ .
\ee  
The second law specifies how the state evolves when an `ideal measurement' is carried out on the subsystem at time $t_m$. It is
\be  |\psih(t_m)\rangle  \to  \frac{s |\psih(t_m)\rangle}{||\, s|\psih(t_m)\rangle\, ||}\quad  \ \  \text{(II)} 
\label{reduction}
\ee
where $s$ is the projection onto the measurement outcome\footnote{von Neumann called I and II,  2 and 1 respectively, but today the given covention is more used.}. 

The Schr\"odinger equation (I) is time reversible --- it can be run both forward and backward in time. The second law of evolution (II) is not reversible. It operates only forward in time. That defines  the quantum mechanical arrow of time. 

It is commonly assumed that the quantum arrow of time coincides with the thermodynamic arrow defined by the direction in which total entropy is increasing.  This identification of a fundamental quantum arrow with a classical one must have seemed natural in a theory which posited separate classical and quantum worlds with a kind of movable boundary between them. A second law describing an ``irreversible act of amplification'' from the quantum world to the classical one in a measurement  was naturally connected with the second law of thermodynamics describing more general classical irreversible processes. 

A thermodynamic arrow of time is not an inevitable feature of a classical world like ours governed by time-neutral dynamical laws. The fact that presently isolated subsystems are mostly evolving towards equilibrium in the same direction in time cannot be a consequence of time-neutral dynamical laws. Rather our thermodynamic arrow arises because the initial state of our universe is such that the progenitors of today's isolated subsystems were all far out of equilibrium a long time ago. As Boltzmann wrote over a century ago: ``The second law of thermodynamics can be proved from the
[time-reversible] mechanical theory, if one assumes that the present state of the
universe\dots started to evolve from an improbable [i.e. special] state''  \cite{Bol97}. Our thermodynamic arrow of time is an emergent feature of the particular initial condition of our universe. 

%Is there  quantum arrow of time even without a thermodynamic arrow?  Or, is our quantum arrow also an emergent feature of the conditions that specify our particular universe in  a more general time-neutral formulation of quantum mechanics. If so, is the quantum arrow always aligned with the thermodynamic arrow or can the two  point in different directions? Such questions are the subject of this paper. To answer them, as Boltzmann noted, we are naturally led to cosmology. 

Is the quantum mechanical arrow of time a fundamental property of quantum mechanics or can it  also be seen as an emergent feature of our universe in a more general formulation of quantum mechanics free from arrows of time? Is a quantum mechanical arrow of time always codirectional with a thermodynamic arrow? Do arrows always point in one direction over the whole of spacetime or can they point in different directions in different spacetime regions? Such questions are the subject of this essay. To answer them, as Boltzmann noted, we are naturally led to cosmology.

\begin{figure}[t]
\includegraphics[height=3in]{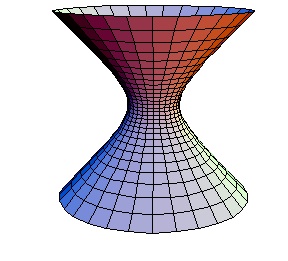}\hfill
\caption{A bouncing universe. The figure shows the geometry of a two-dimensional slice of a four-dimensional cosmological bouncing spacetime embedded in a Lorentz signatured three-dimensional flat space. Time is up. The universe is spatially closed. It has a large spatial volume at large negative times, collapses to smaller and smaller volumes until a minimum is reached (the bounce), and then expands to  larger volumes at positive times. For simplicity we have assumed that the contraction and expansion are time  symmetric. DeSitter space is a well known example. }
\label{fig1}
\end{figure}

 It is almost certain that there will not be a thermodynamic arrow of time that points consistently in one direction over the whole of spacetime in the vast universes contemplated by contemporary inflationary cosmology. Rather the thermodynamic arrow may point in different directions in different regions of spacetime. A simple example that we will discuss in this paper is illustrated in Figure \ref{fig1}.
 
%%%%%%%%%%%%%%%%%%%%%%%%%%%%%%%%%%%%%%%%%%%%%%%%%%%%%
%%%%%%%%%%%%%%%%%%%%%%%%%%%%%%%%%%%%%%%%%%%%%%%%%%%%%%%

The figure shows a  curved two-dimensional slice of a four-dimensional homogeneous and isotropic cosmological spacetime embedded in a (Lorentz signatured) three-dimensional flat space. The universe begins at large radii at the bottom, contracts to smaller and smaller radii, bounces at a minimum radius and reexpands towards the top. DeSitter space is a classic example of such a bouncing spacetime.  

Such bouncing geometries are among the classical spacetimes predicted by the no-boundary quantum state of the universe (NBWF) in simple models using a framework that includes quantum gravity \cite{HHH08, HH12}.  In addition to predicting the behavior of homogeneous and isotropic classical backgrounds the NBWF also predicts the behavior of the quantum fluctuations in matter and geometry away from these symmetries. The key result for the present discussion is the following: The NBWF predicts that fluctuations are in a state of low excitation (ie small)  at the bounce. On either side they grow away from the bounce, become classical,  collapse gravitationally, and eventually create a large scale structure of galaxies, stars, and planets. The thermodynamic arrow of time therefore points in opposite directions on opposite sides of the bounce. The universe has two spacetime  regions with opposing thermodynamic arrows. We can then ask:  What are the laws of quantum mechanics for the quantum fluctuations? Do they have a quantum arrow of time, and  which way does it point in the two regions? 

We will answer these questions in the context of a family of generalizations of textbook quantum theory that are time-neutral --- not preferring one time direction over another and without any built in quantum arrows of time.
In a pioneering paper Aharonov, Bergmann, and Lebovitz \cite{ABL64} showed how to use initial and final conditions to construct such a time-neutral quantum mechanics of measured subsystems. We will answer these questions, not in this context, but in the more general time-neutral quantum mechanics of closed systems that are suitable for cosmology \cite{GH93b}.

In a time-neutral formulation no arrows of any kind are built in. Arrows of time of a particular universe emerge from the conditions that specify it. As we have discussed, the thermodynamic arrow of time in our universe emerges from a special initial condition (and a final condition of indifference.) We will show that quantum mechanical arrows of time can emerge in a similar way. 

For the simple example illustrated in Figure \ref{fig1}, we will find that in a suitable generalization of quantum mechanics there is both a quantum and thermodynamic arrow of time pointing away from the bounce on each side. The overall situation is time symmetric.

More generally we will conclude that quantum mechanical arrows of time are not an inevitable feature of quantum mechanics. Quantum mechanics can be formulated without them. The arrows of time that characterize the approximate quantum mechanics of measured subsystems obeying laws I and II in our universe arise in particular spacetime regions from the conditions that specify the universe and the region. 

The paper is organized as follows:  In Section \ref{ABL} we review the work of Aharonov, Bergmann, and Lebovitz \cite{ABL64}.  Section \ref{qmclosed} introduces a time-neutral quantum mechanics of closed systems with initial and final conditions. All arrows of time arise from asymmetries between these  two.  Section \ref{genqm} introduces the class of generalized quantum theories of which the one in Section \ref{qmclosed} is but one example. Section \ref{bouncing} constructs a time-neutral generalized quantum theory for the quantum fluctuations in a bouncing universe illustrated in Figure \ref{fig1}. Section \ref{conclusions} draws some brief conclusions. 

%%%%%222222222222222222222222222222222222222222222222222222222222
\section{Time-neutrality in  the quantum mechanics of measured subsystems}
\label{ABL}
In a seminal paper Aharonov, Bergmann, and Lebovitz \cite{ABL64} showed how the quantum mechanics of measured subsystems could be formulated without an intrinsic arrow of time by allowing for final states as well as initial ones. We summarize the essence of their argument here in a notation that is analogous to that we will use for cosmology in subsequent sections.

Consider a subsystem of the universe whose states are vectors in a Hilbert space $\Hc_{\rm sub}$.  Alternatives at a moment of time can be reduced to a set of yes/no questions. For instance asking for the position of a particle is equivalent to taking an exhaustive set of position intervals and asking whether the particle is in the first interval (yes or no), then the second interval (yes or no), etc.  In the Heisenberg picture such yes/no alternatives at a time $t$ are represented by an exhaustive set of exclusive projection operators  $\{s_\alpha(t)\}$. For instance, the these might be projections onto an exhaustive set of ranges of position as discussed above. These operators satisfy
\be
\label{projections}
s_\alpha(t)s_{\alpha'}(t) = \delta_{\alpha\alpha'} s_\alpha(t), \quad \sum_\alpha s_\alpha(t) = I . 
\ee
showing that they are projections, that they are exclusive, and that they are exhaustive. 
The projection operators representing the same alternative at two different times $t_1$ and $t_2$ are  connected by the Heisenberg equations of motion
\be
s_\alpha(t_2) = e^{+ih(t_2-t_1)/\hbar} s_\alpha(t_1) e^{-ih(t_2-t_1)]\hbar} .
\label{heiseq}
\ee
where $h$ is the Hamiltonian of the subsystem in isolation.

Suppose a sequence of ideal measurements\footnote{An ideal measurement, sometimes called a projective measurement, is one that disturbs the subsystem as little as possible so that after the measurement its  (Schr\"odinger picture) state is given by \eqref{reduction}.} is carried out on the subsystem by another subsystem at a sequence of times $t_1,t_2, \cdots t_n$. The measurements are described by a sequence of {\it sets} of projections 
$\{s^k_{\alpha_k}(t_k)\}$, $k=1,2,\cdots n$. The upper index allows for the contingency that the measurements might be of different quantities at different times --- a measurement of  position at time $t_1$, of momentum at time $t_2$, etc. 

Suppose that the subsystem is in a (Heisenberg picture) state $|\psi_i\rangle$ in $\Hc$. Then the joint probability for a history of outcomes $\alpha_1,\alpha_2, \cdots \alpha_n$ is \cite{Gro52}
\be
p(\alpha_n, \cdots, \alpha_1) = ||s^n_{\alpha_n}(t_n) \dots s^1_{\alpha_1}(t_1)|\psi_i\rangle||^2 \ .
\label{probhist}
\ee

It is easy to work out that this compact formula for the joint probability for a sequence of  ideal measurement outcomes follows from the two laws of evolution \eqref{seqn} and \eqref{reduction} --- evolve, reduce, evolve, reduce, evolve ...  The formula can be made more compact by defining $\alpha \equiv (\alpha_1,\alpha_2, \cdots \alpha_n)$ and 
\be
c_\alpha \equiv s^n_{\alpha_n}(t_n) \dots s^1_{\alpha_1}(t_1) .
\label{class-meas}
\ee
Then
\be
p(\alpha) = ||c_\alpha|\psi_i\rangle||^2 = || s^n_{\alpha_n}(t_n) \dots s^1_{\alpha_1}(t_1) |\psi_i\rangle||^2 .
\label{prob-c}
\ee
A quantum mechanical arrow of time is manifest in \eqref{probhist} and \eqref{prob-c}. On one end of the chain of projections there is the state, and on the other end there is nothing\footnote{The quantum mechanical arrow of time does not arise from the time-ordering of the projections. That could be reversed by a CPT transformation since field theory is invariant under $CPT$.   But there would still be the state on one end and nothing on the other.}. 
Aharonov, Bergmann, and Lebovitz noticed that if one added a final state $\psi_f$ corresponding to post-selection then the formula for the probabilities becomes
\be
p(\alpha)= N |\langle\psi_f | c_\alpha|\psi_i\rangle|^2, \quad N\equiv|\langle\psi_f |\psi_i\rangle|^{-2}.
\label{prob-meas}
\ee
These formulae are symmetric in the initial and final states, in particular one can write \eqref{prob-meas} as
\be
p(\alpha)= N |\langle\psi_f | c_\alpha|\psi_i\rangle|^2 = N |\langle\psi_i| c^\dagger_\alpha|\psi_f\rangle|^2 .
\label{prob-conj}
\ee
That is, the probabilities are the same when the order of the the projections is reversed and the notion of initial and final interchanged.  

From this perspective, the quantum mechanical arrow of time arises from not specifying a final state. As Aharonov and Rohrlich say \cite{AR05}, ``By imposing an initial and not a final condition we have already sent the arrow of time flying.'' 

\section{A Time-neutral formulation of the quantum mechanics of closed systems.}
\label{qmclosed}

\subsection{A Model Quantum Universe}
\label{model}
Cosmology provides not only the most general context for a discussion of arrows of time but also the most relevant one.  That is because the observed arrows operate on cosmological scales and can be explained by cosmological conditions. 
For instance, as far as we know, the thermodynamic arrow of time extends over the whole of the visible universe and holds from the time of the big bang to the distant future. The evidence of the observations is that the universe was more ordered earlier than now and that disorder has been increasing ever since \cite{BL90,Zhao12}. That is the thermodynamic arrow of time. Similarly the electromagnetic arrow --- the retardation of electromagnetic radiation --- arises because the early universe has very little free electromagnetic radiation that today would be at readily accessible wavelengths \cite{Har05b}.   
The psychological arrow of time can be seen to follow from the other two \cite{Har05b}.

\begin{figure}[t]
\includegraphics[width=3in]{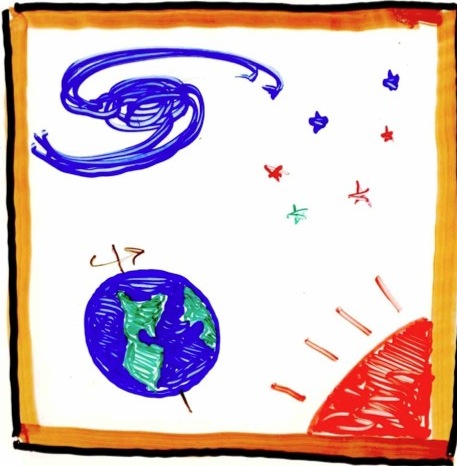}\hfill
\caption{A simple model of a closed quantum system is a universe of quantum matter fields inside a large closed box (say, 20,000 Mpc on a side) with fixed flat spacetime inside. Everything is a physical system inside the box --- galaxies, stars, planets, human beings, observers and observed, measured and measuring. The most general objectives for prediction are the probabilities of the individual members of decoherent sets of alternative coarse grained histories that describe what goes on in the box. That includes histories describing any measurements that take place there. There is no observation or other intervention from outside.  }
\label{box}
\end{figure}

To keep the discussion manageable, we consider a closed quantum system in the approximation that gross quantum fluctuations in the geometry of spacetime
can be neglected\footnote{For the further generalizations that are needed for quantum spacetime see
{\it e.g.} \cite{Har95c,Har07}. For discussions of the arrows of time in contexts that include quantum spacetime see \cite{HH12} and the earlier references therein especially \cite{HLL93}.}. closed system can then be thought of as a large (say
$\gtrsim$ 20,000 Mpc), perhaps expanding, box of particles and fields in a 
fixed, flat, background spacetime (Figure \ref{box}).
Everything is contained within the box, in particular galaxies, planets, observers and
observed, measured subsystems, and any apparatus that measures
them.  This is the most general physical context for prediction.

%%%%%%%%%%%%%%%%%%%%%%%%%%%%%%%%%%%%%%%%%%%%%%%%%%%%%
%%%%%%%%%%%%%%%%%%%%%%%%%%%%%%%%%%%%%%%%%%%%%%%%%%%%%%%

\subsection{Time-Neutral Decoherent Histories Quantum Theory}
\label{DH}
The quantum mechanics of this model universe is formulated in a Hilbert space $\Hc$ that is vastly larger than the Hilbert space of any isolated subsystem it contains. However, the kinematics of the prediction of probabilities for histories bears many similarities to the quantum mechanics of measured subsystems as presented in  the previous section. 

The most general objective of a quantum mechanics of the universe is the prediction of the probabilities for sets of alternative coarse-grained  time histories of its contents. Alternatives at one moment of time are described by an exhaustive set of exclusive projection operators $\{P_\alpha(t)\}$ acting in $\Hc$. These satisfy [cf \eqref{projections}] 
\be
\label{projections1}
P_\alpha(t)P_{\alpha'}(t) = \delta_{\alpha\alpha'} P_\alpha(t), \quad \sum_\alpha P_\alpha(t) = I . 
\ee

A set of alternative coarse-grained histories is specified by a sequence of such sets at a series of times $t_1,t_2, \cdots t_n$. An individual history corresponds to a particular sequence of events  $\alpha \equiv (\alpha_1,\alpha_2, \cdots, \alpha_n)$ and is represented by the corresponding chain of projections:
\be
C_\alpha \equiv P^n_{\alpha_n}(t_n) \dots P^1_{\alpha_1}(t_1) .
\label{class-closed}
\ee
An immediate consequence of this and \eqref{projections1} is that 
\be
\sum_\alpha C_\alpha =I \ ,
\label{exhaustC}
\ee
showing that the set of histories is exhaustive.

This description of histories is  analogous to those in the quantum mechanics of measured subsystems [cf \eqref{class-meas}]. However, there are at least two crucial differences. First, there is no posited separate classical world. It's all quantum. Second, the alternatives represented by the $P$'s are not restricted to measurement outcomes. They might, for example, refer to the orbit to the Moon when no one is looking at it, or to the magnitude of density  fluctuations in the early universe when there were neither observers nor apparatus to measure them. Laboratory measurements can of course be described in terms of correlations between two particular kinds of  subsystems of the universe --- one being measured the other doing the measuring. But laboratory measurements play no central role in formulating the theory, and are just a small part of what it can predict\footnote{Indeed, the quantum mechanics of measured subsystems in Section \ref{ABL} is an approximation appropriate for measurement situations to the more general quantum mechanics of closed systems. See, e.g. \cite{Har91a} Section II.10.}. 

A time-neutral decoherent histories quantum mechanics of our model universe with both initial and final conditions was formulated by Gell-Mann and the author in \cite{GH93b}. The formula for the probabilities for histories is 
\be 
\label{prob-closed}
p(\alpha)= N Tr(\rho_f C_\alpha \rho_i C_\alpha^\dagger) ,  \quad N^{-1}\equiv Tr(\rho_f \rho_i) \ .
\ee
Here, $\rho_i$ and $\rho_f$ are density matrices representing initial and final conditions. There is a clear analogy with \eqref{prob-meas}. 
This expression is time-neutral because initial and final density matrices can be interchanged and the order of times in the $C_\alpha$'s reversed using the cyclic property of the trace. 

\begin{figure}[t]
\includegraphics[height=2in]{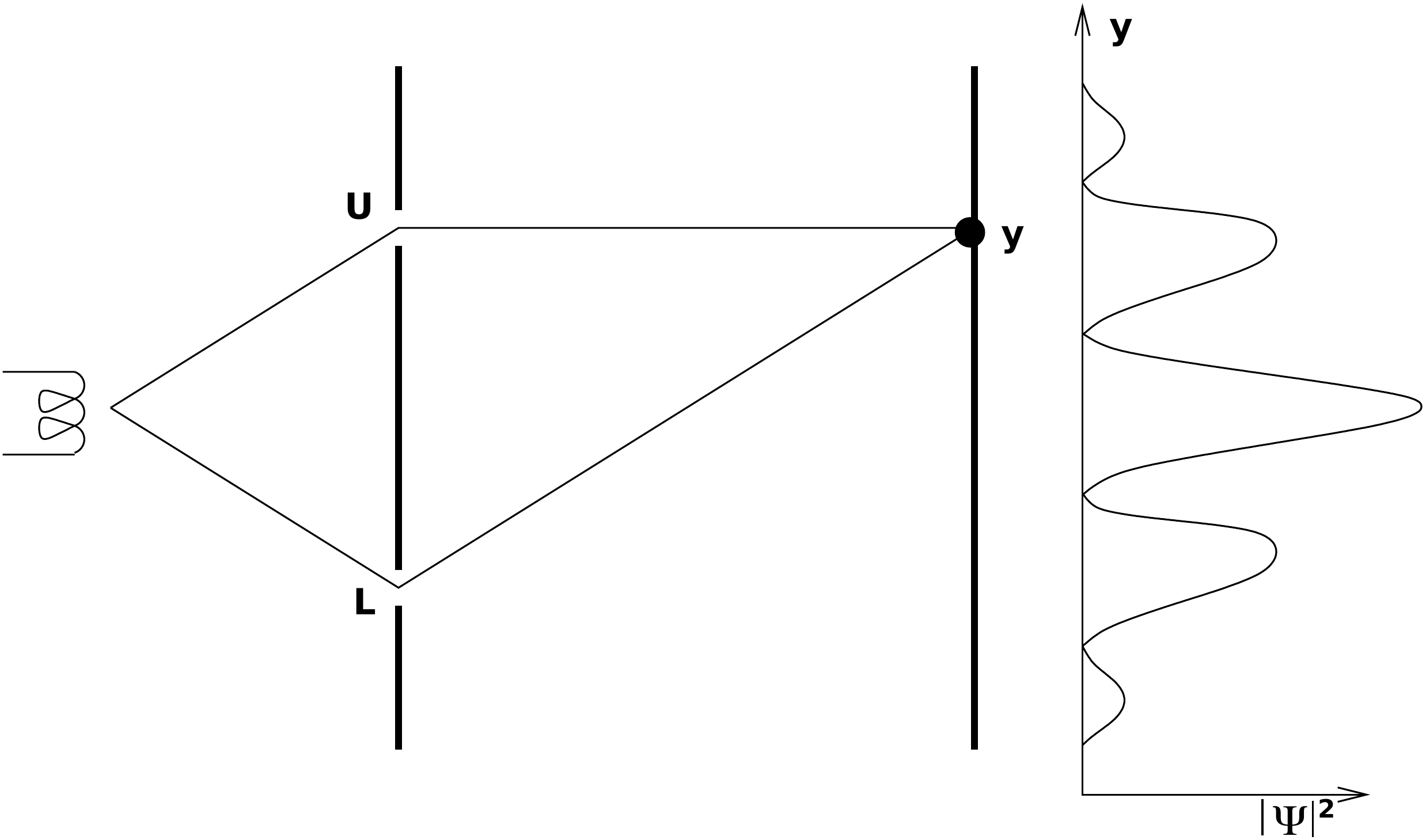}
\caption{The two-slit experiment.  An
electron gun at left emits an electron which is detected at a point $y$ on a screen after passing through another screen with two slits. Because of quantum interference, it is not possible to assign probabilities to the alternative histories in which the electron arrives at $y$  having gone through the upper or lower slit. The probability to arrive at $y$ should be the sum of the probabilities of the two histories.  But in quantum mechanics probabilities are squares of amplitudes and  $|\psi_L (y) + \psi_U (y) |^2 \not= |\psi_L (y) |^2 + |\psi_U (y) |^2$. In a different physical situation where the electron interacts with apparatus that
measures which slit it passed through, then quantum interference is destroyed and consistent probabilities can be assigned to these histories. }

\label{fig2}
\end{figure}

However, \eqref{prob-closed} does not supply probabilities for  all sets of alternative histories. The resulting probabilities might not be consistent with the usual sum rules of probability theory. Generally probabilities cannot be assigned to interfering alternatives in quantum theory. The two-slit experiment described in Figure \ref{fig2} is a simple example.

%%%%%%%%%%%%%%%%%%%%%%%%%%%%%%%%%%%%%%%%%%%%%%%%%%%%%
%%%%%%%%%%%%%%%%%%%%%%%%%%%%%%%%%%%%%%%%%%%%%%%%%%%%%%%

In decoherent (or consistent) histories quantum theory probabilities are assigned only a set of alternative histories if the quantum interference between members of the set is negligible as a consequence of the initial and final conditions and the dynamics. The measure of quantum interference is provided by the decoherence functional:
\be
\label{Dfnal}
D({\alpha},{\alpha'})= N Tr(\rho_f C_{\alpha} \rho_i C_{\alpha'}{^\dagger})
\ee
where $N^{-1}=Tr(\rho_f\rho_i)$. 
A set decoheres when the off diagonal elements of $D$ are negligible. The diagonal elements then give probabilities \eqref{prob-closed} that are consistent with all the rules of probability theory\footnote{For more complete expositions of  decoherent histories quantum theory than the brief synopsis  given here the reader can consult the classic expositions \cite{Gri02,Omn94,Gel94}, a short tutorial in \cite{Har93a}, or a review in \cite{Hoh10}. }. Like \eqref{prob-closed}, the decoherence functional  \eqref{Dfnal} is time neutral. 

In the quantum mechanics of closed systems decoherence replaces `measured' as the criterion for when a set of histories can be consistently assigned probabilities. Measured histories decohere, but histories do not have to be of measurement outcomes in order to decohere. Decoherence is a more precise, more general, and more objective criterion than `measured' and certainly more useful in cosmology.

\subsection{Emergent Arrows of Time}
\label{arrows}
As already mentioned, the expressions both for probabilities \eqref{prob-closed} and interference \eqref{Dfnal} are time-neutral. There is thus no distinction between `initial' and `final' that is not conventional. This formulation of quantum theory for our model universe is therefore free from any built in arrow of time.

If there is no arrow of time in the basic formulation of quantum theory, then the observed arrows of time observed in our particular universe can only arise from differences between the $\rho_f$ and $\rho_i$ that characterize it. We will then say that arrows of time {\it emerge} for our particular universe from $\rho_f$ and $\rho_i$. We will discuss only quantum arrows and thermodynamic arrows, since, as already mentioned, other arrows are connected to these. 

Our observations of the universe from laboratory to cosmological scales are consistent (so far) with one special condition that might be a pure state $\rho_i =|\Psi\rangle\langle \Psi |$ and a second condition of indifference\footnote{For some discussions of the observable information about the final condition, see, e.g. \cite{GH93b,Laf93,Cra95,Pri96}.} $\rho_f \propto I$. It is conventional to call the special condition `initial', as we have done here, the second one `final', and define the direction of increasing time from initial to final.   

With these initial and final conditions the formula for the decoherence functional defining quantum mechanics in the box becomes 
\be
\label{dfnal-usual}
D(\alpha, \alpha') = Tr(C_\alpha \rho_i C^\dagger_{\alpha'}) .
\ee
In particular the probabilities for the histories in a decoherent set are:
\be
p(\alpha_n, \cdots, \alpha_1) =||C_\alpha |\Psi\rangle||^2 =  ||P^n_{\alpha_n}(t_n) \dots P^1_{\alpha_1}(t_1)|\Psi\rangle||^2 .
\label{probhist-closed}
\ee
This has a state on one end of the chain and nothing on the other. Thus, for $\rho_i =|\Psi\rangle\langle \Psi |$ and $\rho_f \propto I$ a quantum mechanical arrow of time emerges [cf.  \eqref{prob-c}].  It is not an arrow that is associated just with histories of measurement situations but more generally with any set of alternative histories of the universe. 

With further assumptions on $\rho_i$ we also recover the thermodynamic arrow.  Suppose the usual entropy of chemistry and physics\footnote{Entropy depends on coarse graining. The usual entropy is defined in terms of a coarse graining expressed in the variables that occur in the deterministic equations of classical physics like the Navier-Stokes equation. For more on its construction and its relation to the quasiclassical realms that are features of our universe see, e.g. \cite{GH07}.}   is low for $\rho_i$. It will be maximal for $\rho_f \propto I$.  It will therefore tend to increase from the time of the initial condition to that of the final one. That is the simplest characterization of the thermodynamic arrow.

However if both $\rho_i$ and $\rho_f$ are non-trivial then there is generally no clear definition of either a global thermodynamic or quantum mechanical arrow. For instance when  $\rho_i$ and $\rho_f$  have comparable low entropies classical analyses \cite{Coc67,GH93b} suggests that the entropy could first rise, and then after a time decrease, leading to a thermodynamic arrow that is local in time first pointing one way and then another\footnote{There will also generally not be a notion of state at a moment of time. However there might be a way of expressing the probabilities in terms of two state vectors similarly to \cite{AV08}. }. 

There is  no clear meaning to a  local quantum mechanical arrow but also no physical need for one. With non-trivial initial and final $\rho$'s there is no notion of a single state at a moment of time from which either the future or the past could be predicted \cite{GH93b}. The theory is fully four-dimensional\footnote{Advanced civilizations with large laboratories and enough money could in principle reverse the thermodynamic arrow of time over a region of spacetime in a universe like ours by pre- and post- selection of quantum states. If they selected initial states at one time indifferently, and states  at a later time distributed according to a low entropy set of probabilities, they would have effectively have reversed both the quantum mechanical and thermodynamic arrows (e.g. \cite{Har91a}, Figure 8). }.

%%%%%33333333333333333333333333333333333333333333333333333333333333333333333333333
\section{Generalized Quantum Theory}
\label{genqm}

The time-neutral formulation of quantum mechanics in the previous section is as notable for its simplicity as it is for its freedom from a built in quantum arrow of time. Formulating quantum theory has been reduced to just two specifications: (1) The sets of possible alternative coarse-grained histories $\{C_\alpha\}$, and (2) a decoherence functional \eqref{Dfnal} that  measures the quantum interference between histories and specified their probabilities when the set decoheres. 

The decoherence functional of time-neutral quantum mechanics  \eqref{Dfnal} is a generalization of that for usual quantum mechanics \eqref{dfnal-usual}.  But it is not the only generalization.  The essential features of quantum mechanics are captured by any complex valued decoherence functional $D(\alpha,\alpha')$ that satisfies the following conditions \cite{Har95c,Ish94}:
\renewcommand{\theenumi}{\roman{enumi}}
\begin{enumerate}

\item{} Hermiticity:  $D(\alpha,\alpha') = D^*(\alpha',\alpha)$ ,

\item{} Normalization:  $\sum_{\alpha\alpha'} D(\alpha,\alpha') =1 $ , 
 
\item{} Positivity:  $D(\alpha,\alpha) \ge 0$ ,

\end{enumerate}
and, most importantly,  consistency with the principle of superposition. This means the following:  Partitioning a set of histories $\{C_\alpha\}$ into bigger sets $\{C_{\bar\alpha}\}$ is an operation of coarse graining. Every history $C_\alpha$ is in one  and only one of the sets $C_{\bar\alpha}$  a fact that we indicate schematically by $\alpha\in \bar\alpha$. Then consistency with the principle of superposition means\footnote{The is just the usual superposition of amplitudes applied to a quantity $D$ that is bilinear in amplitudes.}:    
\vskip .2in                                                                                                                                                                                                            
\indent iv. \ Principle of superposition:  \\ \mbox{} \quad\quad\quad  $D(\bar\alpha,\bar\alpha') = \sum_{\alpha\in\bar\alpha}\sum_{\alpha'\in\bar\alpha'}D(\alpha,\alpha')$ .
\vskip .2in 
Given a decoherence functional satisfying i-iv, the central formula of quantum mechanics which specifies both which sets of histories $\{C_\alpha\}$ decohere and their probabilities $p(\alpha)$ is:
\be
D(\alpha,\alpha') \approx \delta_{\alpha\alpha'}p(\alpha) .
\label{centformula}
\ee
Interference between histories vanishes when the decoherence functional is diagonal and the diagonal elements are the probabilities of the histories in a decoherent set. These probabilities satisfy all the usual rules of probability theory as a consequence of i-iv. 

The decoherence functional of usual quantum mechanics \eqref{dfnal-usual} satisfies i - iv. All the other ways of satisfying these conditions give generalizations of usual quantum theory --- generalized quantum mechanics. The decoherence functional \eqref{Dfnal} of the time-neutral formulation is one example. We will  see another in the next section. 

\section{Bouncing Universe Models}
\label{bouncing}
The universe of quantum matter fields in a closed box that has been at the center of our discussion so far is a much oversimplified model for cosmology. It's chief deficiency is that it ignores gravity. A better, still manageable, kind of model describes quantum matter fields moving in a fixed, classical, cosmological background such as the bouncing universe shown in Figure \ref{fig1}.  What kind of decoherence functional should we assume for such a model to study classical and quantum arrows of time? A model problem in quantum cosmology suggests the answer. 

In \cite{HHH08} both spacetime geometry and matter fields were treated quantum mechanically. Probabilities for different homogeneous and isotropic classical background spacetimes and the behavior of quantum matter fields in them were predicted from the no-boundary quantum state of the universe \cite{HH83} in a simple minisuperspace model. There were two key results for the present discussion. (1) Some bouncing classical background spacetimes like that in Fig \ref{fig1} were predicted with non-zero probability\footnote{Backgrounds that are not time symmetric were also predicted but for simplicity we are focussing on time symmetric ones.}. (2) The predicted matter fields were small and simple at the bounce where the spatial volume is the least, and  not at one or the other of the infinite volume ends of the  spacetime  \cite{HH12}. That suggests that the quantum mechanics of matter fields in such spacetimes should not have initial and final density matrices but rather one density matrix $\rho_0$ at the bounce. We now produce a generalized quantum theory with this property. 

Before starting on quantum mechanics it is worthwhile to consider the thermodynamic arrow of time in this model\footnote{See \cite{HH12} for a more detailed discussion within quantum cosmology and also \cite{HN64,CC04} for not unrelated ones outside of quantum cosmology. }. As discussed above, the matter field fluctuations are small near the bounce. They will therefore grow in the two time directions away from the bounce. Eventually fluctuations may grow large enough to collapse and dissipate giving rise to a large scale structure of galaxies, stars, planets, biota, IGUSes, civilizations, etc on both sides of the bounce\footnote{This large scale structure will generally not be the {\it same} on both sides of the bounce. Individual histories do not have to be time-symmetric. It is the ensemble of possible histories predicted by quantum mechanics  that is time symmetric \cite{HLL93,HH12}. }. The thermodynamic arrow of time is thus bidirectional in this model --- pointing away from the bounce on both sides.

Generalized quantum mechanics for quantum fields in a bouncing universe an be constructed by specifying first the histories and then a decoherence functional obeying properties i-iv in the previous section.  There will be many ways of doing this like simply generalizing the time-neutral formulation of Section \ref{DH} with initial and final density matrices at the large ends of the expansion. But motivated by the quantum cosmology model described above, we are looking for a decoherence functional with a density matrix at the bounce\footnote{We should stress that we are not deriving this decoherence functional from the more general quantum cosmological model that includes quantum spacetime, but using that as a motivation to posit a particular kind of model.}.

%%%%%%%%%%%%%%%%%%%%%%%%%%%%%%%%%%%%%%%%%%%%%%%%%%%%%
\begin{figure}[t]
\includegraphics[width=3in]{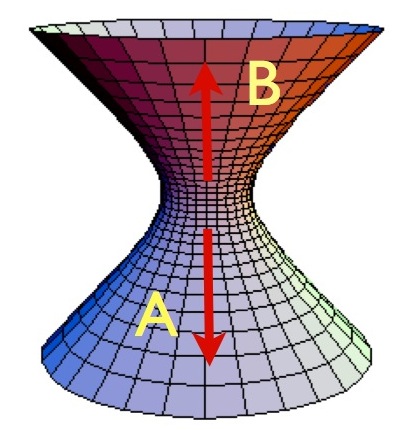}
\caption{A bouncing universe  like that described in Figure \ref{fig1} divided into two sides $A$ and $B$ at the minimum volume three surface (the bounce).  When the  quantum mechanics  of matter fields is described by a decoherence functional \eqref{df-bounce} each side has a coincident quantum and thermodynamic arrow. But the arrows point in opposite directions on opposite sides of the bounce.  This is an example whose quantum mechanics is globally time neutral but with local arrows of time. }
\label{halves}
\end{figure}
%%%%%%%%%%%%%%%%%%%%%%%%%%%%%%%%%%%%%%%%%%%%%%%%%%%%%%%

To specify the histories we arbitrarily label the two sides of the bounce as $A$ and $B$ as in Figure \ref{halves}. A given history will generally have a part on the $A$ side and a part on the $B$ side --- generally different.  On each side the parts of  histories can be represented by chains of projections --- $\{C^A_\alpha\}$ and $\{C^B_\beta\}$. We make the convention that the projections in the chains are {\it time-ordered away from the bounce}.  We have separately [cf \eqref{exhaustC}]
\be
\label{exhaustab}
\sum_\alpha C^A_\alpha =I, \quad   \sum_\beta C^B_\beta =I \ .
\ee

The following decoherence functional then suggests itself
\be
\label{df-bounce}
D(\beta,\alpha;\beta',\alpha') = Tr( C^B_{\beta} \sqrt{\rho_0} C^{A\dagger}_{\alpha} C^{A}_{\alpha'} \sqrt{\rho_0} C^{B\dagger}_{\beta'}) \  .
\ee
It is not difficult to verify that this satisfies requirements i-iv. 

The generalized quantum theory defined by \eqref{df-bounce} is time neutral. The decoherence functional $D$ is symmetric under interchanging $A$ and $B$. It is perhaps the simplest generalized quantum theory with this property\footnote{It is not, however, the only one. For instance initial and final conditions represented by density matrices $\rho_i$ and $\rho_f$ at the large ends could have been incorporated in addition to $\rho_0$ in analogy with \eqref{Dfnal}.}. 

Familiar results emerge if we consider histories just on one side, say $B$. The appropriate decoherence functional $D^B$  results from coarse-graining (summing) over alternatives on the $A$ side. Using \eqref{exhaustab} we find 
\be
D^B(\beta, \beta') = Tr(C_\beta \rho_0 C^\dagger_{\beta'}) .
\label{dfB}
\ee
But this is just the expression \eqref{dfnal-usual}.  There will thus be a quantum mechanical arrow of time on side $B$ and a coincident thermodynamic arrow. Similar results are obtained by following histories on side $A$ and ignoring those on side $B$.

Thus  the generalized quantum theory defined by the decoherence functionsl \eqref{df-bounce} exhibits {\it local} thermodynamic and quantum mechanical arrows that are codirectional on either side but point in opposite directions on opposite sides of the bounce. There are no global arrows --- either quantum mechanical or thermodynamic --- pointing consistently in one direction over the whole of the spacetime. 

If we live in a bouncing universe  questions naturally arise as to how much present events on our side are influenced by what occurred before the  bounce and what we can infer about events on the far side from observations on our side.  The answers to such questions are contained in the joint probabilities $p(\beta,\alpha)$ for correlations between histories on the far side of the bounce and histories on the far side. 

We can anticipate that it will be difficult to find causal correlations between the two sides because the thermodynamic arrow points in opposite directions on opposite sides of the bounce \cite{HH12}.  The two sides are in each other's pasts as determined by the thermodynamic arrow. There is as much chance of events on the far side of the bounce influencing us, as we have of influencing events in our past by actions taken now.  

In the simple case where the density matrix $\rho_0$ in \eqref{df-bounce} is pure mutual influence is impossible. To see this  write
\be
\rho_0 = |\Psi\rangle\langle\Psi| = \sqrt{\rho_0} .
\label{purerho}
\ee
Then
\be
D(\beta,\alpha;\beta',\alpha') =\langle\Psi |C^{B\dagger}_{\beta'} C^B_\beta|\Psi\rangle \langle\Psi |C^{A\dagger}_{\alpha' }C^A_\alpha|\Psi\rangle .
\label{df-pure}
\ee
The immediate consequence is that the joint probabilities of a decoherent set of histories factors
\be
\label{factor}
p(\beta,\alpha) =p^B(\beta)p^A(\alpha),
\ee
and there is no correlation between events on one side and the events on the other. The far side might as well not exist.

\section{Conclusions}
\label{conclusions}

Cosmology is the natural context for understanding the origin of the arrows of time our our universe. Arrows operate over cosmological distances and can be explained by cosmological conditions. 

No arrows of time need be built into a fundamental formulation of quantum mechanics of closed systems like the universe. Rather quantum mechanics can be formulated time-neutrally. The observed arrows of time are then emergent features of the asymmetries between conditions that specify our particular universe among the possibilities that the time-neutral theory allows.  This general perspective allows a discussion of the different ways arrows to time can be exhibited by different universes specified by different conditions.  In particular it allows a discussion of the connections between arrows that follow from those conditions. 

This essay exhibited a number of cosmological models with different possibilities for the quantum mechanical and thermodynamic arrows in the framework of time-neutral generalized quantum theory in fixed background spacetimes. From these examples we can conclude that a given universe may not exhibit well defined arrows of either kind.  Further, when arrows do emerge they need not consistently point in one direction over the whole of spacetime. Rather they may point in different directions in different regions of spacetime as the bouncing universe model cleanly illustrates. Local arrows can be consistent with global time-symmetry \cite{HH12,CC04}.

 In some examples there was a local thermodynamic arrow defined by the direction of local entropy increase but no obvious quantum mechanical arrow. In all examples, where both arrows were available locally they coincided in direction. (Of course, a few examples do not make a general result.)

From this perspective, other features of quantum mechanics such as states on spacelike surfaces, their unitary evolution and their reduction may also emerge only locally in a more general framework for quantum theory that is fully four-dimensional and time neutral \cite{Har95c}.

%But from a more general perspective there are a  number of difficulties with this connection between arrows. (1) Realistic measurements seldom obey anything as simple as II. (2) Many measurements are irreversible, but irreversibility is not a general requirement for measurement \cite{REVMEAS}. But most importantly, the theremodynamic arrow of time can be in principle be locally reversed --- say by postselection of an ensemble. Does the quantum mechanical arrow of time also reverse?  If so how does it work?

%%%%%-------------------------------------------------------------------------------------------------------------

\renewcommand{\theequation}{\Alph{section}.\arabic{equation}}

%\vskip .3in 
\acknowledgments  

The author thanks Thomas Hertog for many discussions of the arrows of time in quantum cosmology and Murray Gell-Mann for discussions on the quantum mechanics of the universe over many decades. He thanks the Santa Fe Institute for supporting many productive visits there. The this work was supported in part by the National Science Foundation under grant PHY12-05500.

\end{document}